\documentclass[11pt,epsf]{article}
\usepackage{amsmath}
\usepackage{amsfonts}
\usepackage{amssymb}
\usepackage{graphicx}

\newcommand {\dfn} {\stackrel{\Delta} {=}}
\newcommand {\exe} {\stackrel{\cdot} {=}}

\newcommand {\bu} {\mbox{\boldmath $u$}}

\newcommand {\bx} {\mbox{\boldmath $x$}}
\newcommand {\by} {\mbox{\boldmath $y$}}

\newcommand {\bE} {\mbox{\boldmath $E$}}

\newcommand {\bX} {\mbox{\boldmath $X$}}

\newcommand{\calE}{{\cal E}}

\begin{document}
\thispagestyle{empty}
\title{The Random Energy Model in a Magnetic Field and Joint Source--Channel Coding}
\author{Neri Merhav}
\date{}
\maketitle

\begin{center}
Department of Electrical Engineering \\
Technion - Israel Institute of Technology \\
Haifa 32000, ISRAEL \\
\end{center}
\vspace{1.5\baselineskip}
\setlength{\baselineskip}{1.5\baselineskip}

\begin{abstract}
We demonstrate that there is an intimate relationship between the magnetic
properties of Derrida's random energy model (REM) of spin glasses
and the problem of joint source--channel coding in Information Theory.
In particular, typical patterns of erroneously decoded messages in the
coding problem have ``magnetization'' properties that are analogous
to those of the REM in certain phases, 
where the non--uniformity of the distribution of the source
in the coding problem, plays the role of an external magnetic field applied to the REM.
We also relate the ensemble performance (random coding exponents) of joint source--channel
codes to the free energy of the REM in its different phases.\\

\end{abstract}

\section{Introduction}

In the last few decades it has become apparent that many problems in
Information Theory, and coding problems in particular,
can be mapped onto (and interpreted as) analogous problems in the
area of statistical physics of disordered systems, most notably, spin glass models.
Such analogies are useful because physical insights, as well as
statistical mechanical tools and analysis techniques (like the replica method),
can be harnessed in order to advance the knowledge and the
understanding with regard to
the information--theoretic problem under discussion (and conversely, information--theoretic
approaches to problems in physics may sometimes prove useful to physcists as well).
A very small, and by no means exhaustive,
sample of works along this line includes references [1]--[25].

In particular, Sourlas \cite{Sourlas89},\cite{Sourlas94} was the first
to observe that there are strong analogies and parallisms between 
the behavior of ensembles of error correcting codes 
and certain spin glass models with quenched parameters, like 
the $p$--spin glass model and Derrida's random energy model (REM) 
\cite{Derrida80a},\cite{Derrida80b},\cite{Derrida81}
at least as far as the mathematical formalism goes. In particular, the REM
is an especially attractive model to adopt in this context, as it is,
on the one hand, exactly solvable, and on the other hand, rich enough to
exhibit phase transitions. As noted in
\cite[Chap.\ 6]{MM06} and \cite{MU07}, ensembles of error correcting codes
`inherit' these phase transitions from the REM when viewed as physical
systems whose phase diagram is defined in the plane of the coding rate vs.\ decoding temperature.
In \cite{Merhav07} this topic was further investigated and ensemble performance figures
of error correcting codes (random coding exponents) were related to the free energies in the
various phases of the phase diagram.

While the above--described relation takes place between {\it pure} channel coding and 
the REM {\it without} any external magnetic field, 
in this work, we demonstrate that there are also intimate relationships between {\it combined
source/channel coding} and the REM {\it with} such a magnetic field.
In particular, it turns out that typical patterns of erroneously decoded messages in the
source/channel coding problem have ``magnetization'' properties that are analogous
to those of the REM in certain phases,
where the non--uniformity of the distribution of the source
in the joint source--channel coding system, plays the role of an external magnetic field applied
to the spin glass modeled by the REM.
We also relate the ensemble performance (random coding exponents) of joint source--channel
codes to the free energy of the REM in its different phases.

The outline of this paper is as follows. In Section 2, we provide some background,
both on the information theoretic aspect of this work, which is the problem of joint
source channel coding, and the statistical mechanical aspect, which is the REM and its
magnetic properties. In Section 3, we present the phase diagram pertaining to
finite--temperature decoding of an ensemble of joint source--channel codes and characterize the
free energies in the various phases. Finally, in Section 4, we derive random coding
exponents pertaining to this emsemble and demonstrate their relationships to the
free energies.

\section{Background}

In this section, we give some very basic background which will be needed in the sequel.
In Subsection 2.1, we provide a brief overview of Shannon's fundamental coding
theorems, the skeleton of Information Theory: The source coding theorem, the channel
coding theorem, and finally the joint source--channel coding theorem.
In Subsection 2.2, we review a few models of spin glasses, with special emphasis on
the REM.

\subsection{Information Theory}

\subsubsection{Source Coding}

Suppose we wish to compress a sequence of $N$ bits, $(u_1,u_2,\ldots,u_N)$, drawn from 
a stationary memoryless binary source, i.e., each bit is drawn independently,
where $\mbox{Pr}\{u_i=1\}=1-\mbox{Pr}\{u_i=0\}=q$. Shannon's {\it source coding theorem}
(see, e.g., \cite[Chap.\ 5]{CT06}) tells that if we demand that the source sequence
would be perfectly reconstructable from the compressed data, then the best achievable
compression ratio (i.e., the smallest average ratio between the
compressed message length and the original source 
message length -- $N$), at the limit of large $N$, is given by the entropy of the
source, which in the binary memroyless case considered here, is given by:
$$h(q)=-q\log_2q-(1-q)\log_2(1-q).$$
Many practical coding algorithms are known to achieve $h(q)$ asymptotically, e.g., Huffman coding,
Shannon coding, arithmetic coding, and Lempel--Ziv coding, to name a few \cite{CT06}.

\subsubsection{Channel Coding}

Shannon's celebrated 
{\it channel coding theorem} (see, e.g., \cite[Chap.\ 7]{CT06}) is about reliable transmission
of digital information across a noisy channel: 
Suppose we wish to transmit a binary messsage of $k$ bits, indexed by $m$ 
($0\le m\le 2^k-1$),
through a noisy binary symmetric channel, which flips the transmitted bit with
probability $p$ or conveys it unaltered, with probability $1-p$. 
If we wish to convey the message via the channel reliably (i.e., with very
small probability of error), then before
we transmit the message via the channel, we have to encode it, i.e., map it 
in a sophisticated manner into a longer binary message
of length $n$ ($n\ge k$) and then transmit the encoded message $\bx(m)=(x_1(m),\ldots,x_n(m))$.
The ratio $R=k/n$ is called the {\it coding rate}. It measures how efficiently the channel is
used, i.e., how many information bits are conveyed per one channel use.
The corresponding channel output sequence, $\by=(y_1,\ldots,y_n)$
(with some of the bits flipped by the channel), is received at the
decoder. 

The optimum decoder, in the sense of minimum probability of error, 
estimates the message $m$ by the {\it maximum a--posteriori} (MAP) decoder, i.e., it
selects the message $0\le m\le 2^k-1$ 
which maximizes posterior probability given $\by$, that is, $P(m|\by)$, or equivalently, it
maximizes the product $P(m)P(\by|\bx(m))$, where $P(m)$
the prior probability of message $m$ and $P(\by|\bx(m))$ is the conditional probability
of the observed $\by$ given that $\bx(m)$ was transmitted. In the important special case
where all messages are {\it a-priori} equiprobable, that is, $P(m)=2^{-k}$ for all $m$,
the MAP decoding rule boils down to the maximization of $P(\by|\bx(m))$, which
is the {\it maximum likelihood} (ML) decoding rule. 

Channel capacity $C$ is defined as the supremum
of all coding rates $R$ for which there still exist encoders
and decoders which make the probability of error arbitrarly small provided that $n$ 
is large enough (keeping $R=k/n$ fixed). 
Shannon's channel coding theorem provides a formula of the channel capacity, which
in the binary case considered here, is given by
$$C=1-h(p)=1+p\log_2p+(1-p)\log_2(1-p).$$
One of the mainstream efforts in the Information Theory literature has evolved around
devising practical coding and decoding schemes, 
in terms of computational complexity and storage, with rates close to capacity.

\subsubsection{Joint Source--Channel Coding}

Finally, we consider the problem of {\it joint source--channel coding} (see, e.g.,
\cite[Sect.\ 7.13]{CT06}): Suppose we have a binary
memoryless source, as in the first paragraph above, and a binary memoryless channel, as in
the second paragraph above. We assume that by the time that the source generates $N$ symbols,
the channel can transmit $n=N\theta$ bits ($\theta\ge 0$ is fixed). 

A joint source--channel code
maps the source sequence $\bu=(u_1,\ldots,u_N)$ 
of length $N$ into a channel input sequence $\bx(\bu)$ of length $n$. The decoder,
that receives the channel output vector $\by$, estimates $\bu$ either by the
{\it symbol MAP} decoder, which minimizes the symbol error probability (or the bit error probability)
or the {\it word MAP} decoder, which as mentioned earlier, 
minimizes the word error probability. The word MAP decoder works
similarly to the above described MAP decoder for a channel code:
It estimates the source sequence as a whole by seeking the vector $\bu$ that maximizes 
$P(\bu)P(\by|\bx(\bu))$, where $P(\bu)$ is the probability of the
source vector $\bu$.
The symbol MAP decoder, on the other hand, estimates each bit $u_i$ of the source separately by
seeking the symbol $u\in\{0,1\}$ that maximizes 
$\mbox{Pr}\{u_i=u,\by\}=\sum_{\bu:~u_i=u}P(\bu)P(\by|\bx(\bu))$, $i=1,\ldots,N$.

These two decoders can be thought of as two special cases of a more general class of
decoders, referred to as {\it finite--temperature decoders} \cite{Rujan93}. A finite--temperature
decoder estimates the $i$--th symbol $u_i$ by
$$\hat{u}_i=\mbox{argmax}_{u\in\{0,1\}}\sum_{\bu:~u_i=u}[P(\bu)P(\by|\bx(\bu))]^\beta,$$ 
where the
parameter $\beta$ can be thought of as an inverse temperature parameter. The choice $\beta=1$
corresponds to the symbol MAP decoder, whereas $\beta\to\infty$ gives us the word MAP decoder
\cite[Chap.\ 6]{MM06}. 

The joint source--channel coding theorem asserts that a necessary and sufficient
condition for the existence of codes, that for large enough $n$ and $N$ (with $\theta=n/N$ fixed),
$\bu$ can be decoded with aribrarily small probability of error (both wordwise and symbolwise)
is given by
\begin{equation}
\label{relcond}
h(q)\le \theta C.
\end{equation}
One approach to achieve reliable communication, whenever this condition holds,
is to apply separate source coding and channel coding: First compress the
source to essentially $h(q)$ bits per symbol, resulting in a binary compressed message
of length about $Nh(q)=nh(q)/\theta$ bits, as described in the first paragraph above,
and then use a reliable channel code of rate $R=h(q)/\theta \le C$
to convey the compressed message, as described in the second paragraph. The decoder will 
first decode the message by the corresponding channel decoder and then decompress the resulting
message. Another approach is to map $\bu$ directly to a channel input vector $\bx(\bu)$.
It can be shown \cite[Exercise 5.16, p.\ 534]{Gallager68} that by a random selection of a code 
from the uniform ensemble (i.e., by generating each codeword 
$\bx(\bu)$, $\bu\in\{0,1\}^N$, independently by a sequence of $n$ fair coin tosses),
the average probability of error, over this ensemble of codes, tends to zero
as the block length goes to infinity, as long as the above necessary and sufficient condition holds.

\subsection{The REM}

Consider a spin glass with $n$ spins, designated by a binary vector $\sigma=(\sigma_1,\ldots,\sigma_n)$,
$\sigma_i\in\{-1,+1\}$, $i=1,2,\ldots,n$.
The simplest model of this class is 
that of a {\it paramagnetic} solid, namely,
the one where the only effect is that of the external
magnetic field $H$, whereas the effect of interactions is negligible 
(cf.\ \cite[Chap.\ 3]{Mandl71}). Assuming that the spin directions
are all either parallel or antiparallel to the direction of the external magnetic field, the
energy associated with a configuration $\sigma$ is given (in the appropriate units) by:
$$\calE(\sigma)=-H\sum_{i=1}^n\sigma_i,$$
which means (according to the Boltzmann distribution) that each spin is independently
oriented upward (+1) with probability $e^{\beta H}/[2\cosh(\beta H)]$ or
downward (-1) with probability $e^{-\beta H}/[2\cosh(\beta H)]$.
This means that the average (net) magnetic moment is
\begin{equation}
\label{paramagnetization}
m\dfn (+1)\cdot\frac{e^{\beta H}}{2\cosh(\beta H)}+(-1)\cdot
\frac{e^{-\beta H}}{2\cosh(\beta H)}=\tanh(\beta H)
\end{equation}
and so the average internal energy per particle is 
$\bar{E}=-H\tanh(\beta H)$ and the free energy per particle is
$F=-\frac{1}{\beta}\ln[2\cosh(\beta H)]$.

More involved (and more interesting) situations occur, of course, 
when the effect of mutual interactions among
the spins is appreciable. The simplest model that accounts for interactions is
the {\it Ising model}, given by
\begin{equation}
\label{ham}
\calE(\sigma)=-J\sum_{i,j}\sigma_i\sigma_j-H\sum_{i=1}^n\sigma_i
\end{equation}
where the second term is the contribution of the external magnetic field
as before, and the in the first term, pertaining to the interaction,
$J$ describes the intensity of the interaction with the summation
being defined over pairs of neighboring spins (depending on the geometry of the problem).

More general models allow interactions not only with immediate neighbors,
but also with more distant ones, and then there are different strengths of interaction, depending
on the distance between the two spins. In this case, the first 
term is replaced, by the more general form $-\sum_{i,j}J_{ij}\sigma_i\sigma_j$, 
where now the sum can be defined over all possible pairs 
$\{(i,j)\}$.
Here, in addition to the ferromagnetic case, where all $J_{ij} > 0$, and the
antiferromagnetic case, where all $J_{ij} < 0$, there is also a mixed situation where
some $J_{ij}$ are positive and others are negative, which is the case of a {\it spin glass}.
Here, not all spin pairs can be in their preferred mutual position (parallel/antiparallel),
thus the system may be {\it frustrated.}

To model situations of disorder, it is common to model $\{J_{ij}\}$ as random variables (RV's) with,
say, equal probabilities of being positive or negative. For example,
in the Edwards--Anderson (EA) model \cite{EA75},
$\{J_{ij}\}$  are taken to be i.i.d.\ zero--mean Gaussian RV's when $i$ and $j$ are neighbors
and set to zero otherwise. In the Sherrington--Kirkpatrick (SK) model \cite{SK75}, all $\{J_{ij}\}$ are
i.i.d., zero--mean Gaussian RV's. In the {\it $p$--spin--glass model}, the interaction terms
consist of all products of combinations of $p$ 
spins (rather than just pairs) with Gaussian coefficients of the
appropriate scaling (cf.\ e.g., \cite{DG86a}).

In all these models, the system has two levels of randomness: the
randomness of the interaction coefficients and the randomness of the spin configuration given
the interaction coefficients, according to the Boltzmann distribution. However, the two
sets of RV's are normally treated differently. The random coefficients are commonly considered {\it
quenched} RV's,
namely, they are considered fixed in the time scale at which the spin
configuration may vary. This is analogous to the model
of coded communication in a random coding paradigm: A randomly
drawn code should normally be thought of as a quenched
entity, as opposed to the randomness of the source and/or the channel.

\subsubsection{The REM in the Absence of a Magnetic Field}

In \cite{Derrida80a},\cite{Derrida80b},\cite{Derrida81}, Derrida took the above described
idea of randomizing the (parameters of the) Hamiltonian to an extreme, and suggested a model of spin
glass with disorder under which the energy levels $\{\calE(\sigma)\}$ are simply i.i.d.\ RV's, without
any structure in the form of (\ref{ham}) 
or its above--described extensions. It can also be viewed, however,
as the asymptotic behavior of the $p$--spin--glass model when $p\to\infty$ (a limit to be taken after the
limit $n\to\infty$, i.e., $p<<n$) \cite{DG86a}.
In particular,
in the absence of a magnetic field,
the $2^n$ RV's $\{\calE(\sigma)\}$ are taken to be i.i.d., zero--mean Gaussian RV's, all with variance
$nJ^2/2$, where $J$ is a parameter.\footnote{The variance scales linearly with $n$ to match
the behavior of the Hamiltonian (\ref{ham}) with
a limited number of interacting neighbors and random interaction parameters, which has a number of
independent terms that is linear in $n$.} The beauty of the REM is in that on the one hand, it
is very easy to analyze, and on the other hand, it consists of sufficient richness to exhibit
phase transitions.

The basic observation about the REM is that for a typical realization of the
configurational energies $\{\calE(\sigma)\}$, the density of number of configurations with energy
about $E$ (i.e., between $E$ and $E+dE$), $N(E)$, is proportional
(up to sub--exponential terms in $n$) to $2^n\cdot e^{-E^2/(nJ^2)}$, as long as
$|E|\le E_0\dfn nJ\sqrt{\ln 2}$, whereas energy levels outside this range are typically
not populated by spin configurations ($N(E)=0$), as the probability
of having at least one configuration with such an energy decays exponentially with $n$.
Thus, the asymptotic (thermodynamical) entropy per spin, which is defined by
$$S(E)=\lim_{n\to\infty}\frac{\ln N(E)}{n}$$
is given by
$$S(E)=\left\{\begin{array}{ll}
\ln 2 -\left(\frac{E}{nJ}\right)^2 & |E|\le E_0\\
-\infty & |E| > E_0 \end{array}\right.$$
The partition function of a typical realization of a REM spin glass
is then
\begin{eqnarray}
Z(\beta)&\exe&\int_{-E_0}^{E_0}dE\cdot N(E)\cdot e^{-\beta E}\nonumber\\
&\exe&\int_{-E_0}^{E_0}dE \cdot e^{nS(E)}\cdot e^{-\beta E}
\end{eqnarray}
where the notation $\exe$ designates asymptotic equivalence between
two functions of $n$ in the exponential scale.\footnote{More precisely, $a_n\exe b_n$
means that $\lim_{n\to\infty}\frac{1}{n}\log\frac{a_n}{b_n}=0$.}
The exponential growth rate of $Z(\beta)$,
$$\phi(\beta)\dfn\lim_{n\to\infty}\frac{\ln Z(\beta)}{n},$$
behaves according to
\begin{eqnarray}
\phi(\beta)
&=&\max_{|E|\le E_0}\left[S(E)-\beta\cdot\frac{E}{n}\right]\nonumber\\
&=&\max_{|E|\le E_0}\left[\ln 2-
\left(\frac{E}{nJ}\right)^2-\beta J \cdot\left(\frac{E}{nJ}\right)\right].
\end{eqnarray}
Solving this simple optimization problem, one finds that $\phi(\beta)$
is given by
$$\phi(\beta)=\left\{\begin{array}{ll}
\ln 2 +\frac{\beta^2J^2}{4} & \beta \le \frac{2}{J}\sqrt{\ln 2}\\
\beta J\sqrt{\ln 2} & \beta > \frac{2}{J}\sqrt{\ln 2}\end{array}\right.$$
which means that the asymptotic free energy per spin, a.k.a.\ the {\it free energy density},
is given by (cf.\ \cite[Proposition 5.2]{MM06}):
$$F(\beta)=
\left\{\begin{array}{ll}
-\frac{\ln 2}{\beta}-\frac{\beta J^2}{4} & \beta \le \frac{2}{J}\sqrt{\ln 2}\\
-J\sqrt{\ln 2} & \beta > \frac{2}{J}\sqrt{\ln 2}\end{array}\right.$$
Thus, the free energy density is subjected to a phase transition at the inverse
temperature $\beta_c\dfn \frac{2}{J}\sqrt{\ln 2}$. At high temperatures ($\beta < \beta_c$),
which is referred to as the {\it paramagnetic phase}, the partition function is dominated by
an exponential number of configurations with energy $E=-n\beta J^2/2$ and the entropy
grows linearly with $n$. When the system is cooled to $\beta=\beta_c$ and beyond,
which is the {\it glassy phase},
the system freezes but it is still in disorder -- the partition function
is dominated by a subexponential number of configurations of minimum energy $E=-E_0$.
The entropy, in this case, grows sublinearly with $n$, namely the entropy
per spin vanishes, and the free energy density no longer depends on $\beta$.
Further details about the REM can be found in \cite{MM06} and the references mentioned in
the Introduction.

\subsubsection{The REM in the Presence of a Magnetic Field}

The random energy levels of the REM, as described above, represent the interaction energies
among the various spins in the absence of an external magnetic field.
In the presence of an external uniform magnetic field, $H$ (cf.\ 
\cite{Derrida80a},\cite{Derrida80b},\cite{Derrida81}), the Hamiltonian of the system
should be supplemented with the term 
$-H\sum_i\sigma_i=-nm(\sigma)H$ (cf.\ eq.\ (\ref{ham})), where
$$m(\sigma)=\frac{1}{n}\sum_{i=1}^n\sigma_i=
\frac{1}{n}\sum_{i=1}^n[1\{\sigma_i=1\}-1\{\sigma_i=-1\}]=\frac{2n_1(\sigma)}{n}-1$$ 
is the magnetization
associated with the configuration $\sigma$, and $n_1(\sigma)$ is
the number of spins up, $\sum_{i=1}^n1\{\sigma_i=1\}$.
As far as the statistical description of the REM goes, 
this shifts the expectation of the random energy level
$\calE(\sigma)$ from zero to $-nm(\sigma)H$. Equivalently, 
we can assign the same zero--mean Gaussian distribution
as before to the interaction energy, call it now $\calE_I(\sigma)$, 
and add to each configuration $\sigma$ the term
$-nm(\sigma)H$. The corresponding partition function would then be:
\begin{eqnarray}
\label{rempart}
Z(\beta,H)&=&\sum_{\sigma}e^{-\beta[\calE_I(\sigma)-nm(\sigma)H]}\nonumber\\
&=&\sum_m\left[\sum_{\sigma:~m(\sigma)=m}e^{-\beta\calE_I(\sigma)}\right]e^{\beta nmH}\nonumber\\
&\dfn&\sum_m \zeta(\beta,m)e^{\beta nmH}
\end{eqnarray}
where $\zeta(\beta,m)$, referred to as the {\it partial partition function}, contains only the
contributions of configurations whose magnetization 
$m(\sigma)$ is equal to $m$. The behavior of $\zeta(\beta,m)$
is exactly like that of the REM without a magnetic field, except that instead of $2^n$ configurations,
it has only $|\{\sigma:~m(\sigma)=m\}|\exe 2^{nh((1+m)/2)}$ configurations, where $h(\cdot)$ is the binary
entropy function. By carrying out a similar analysis as in the previous subsection
to $\zeta(\beta,m)$ and then finding the dominant contribution of $m$ (which is the typical
magnetization), one can show (cf.\ \cite{Derrida80a},\cite{Derrida80b},\cite{Derrida81}) that there
exists a phase transition at $\beta=\beta_{c}(H)$, where $\beta_{c}(H)$ is the unique solution to
the equation
$$\beta^2J^2=4h\left(\frac{1+\tanh(\beta H)}{2}\right).$$
It is not difficult to see that $\beta_{c}(H)$ is a non--increasing function of $|H|$ and
therefore $T_{c}(H)=1/\beta_c(H)$ is non--decreasing, with a minimum at $H=0$, given by
$T_c(0)=J/(2\sqrt{\ln 2})$ (see Fig.\ 1).
For high temperatures ($\beta \le \beta_c(H)$), 
where the effect of the interactions among the spins is relatively insignificant,
one observes the ordinary paramagnetic behavior,
with the average magnetization is
$$m=m_p(\beta,H)\dfn\tanh(\beta H),$$
whereas for low temperatures ($\beta \ge \beta_c(H)$), the system is frozen in the spin glass phase
where the magnetization no longer depends on the temperature:
$$m=m_g(H)\dfn\tanh(\beta_c(H)\cdot H).$$
The free energy per spin is given by
$$F(\beta,H)=-\lim_{n\to\infty}\frac{\ln Z(\beta,H)}{n\beta}=
\left\{\begin{array}{ll}
-\left[\frac{\beta J^2}{4}+
\frac{h([1+\tanh(\beta H)]/2)}{\beta}+H\tanh(\beta H)\right] & \beta < \beta_c(H)\\
-\left[J\sqrt{h\left(\frac{1+\tanh(\beta_c(H)H)}{2}\right)}+
H\tanh(\beta_c(H)H)\right] & \beta \ge \beta_c(H)
\end{array}\right.$$
As can be seen, no sponteneous magnetization takes place under the REM, even at low
temperatures ($H\to 0$ implies $m\to 0$). 
As for other thermodynamic quantities, we have
the average internal energy per spin
$$E(\beta,H)=\frac{\partial}{\partial \beta}[\beta F(\beta,H)]= 
\left\{\begin{array}{ll}
-H\tanh(\beta H)-\frac{\beta J^2}{2} & \beta < \beta_c(H)\\
-H\tanh(\beta_c(H)\cdot H)-\frac{\beta_c(H)J^2}{2} & \beta \ge \beta_c(H)
\end{array}\right.$$
the entropy per spin
$$S(\beta,H)=\beta[E(\beta,H)-F(\beta,H)]=\left\{\begin{array}{ll}
h\left(\frac{1+\tanh(\beta H)}{2}\right)-\frac{\beta^2 J^2}{4} & \beta < \beta_c(H)\\
0 & \beta \ge \beta_c(H)
\end{array}\right.$$
and the magnetic susceptibility
$$\chi =\left[\frac{\partial m}{\partial H}\right]_{H=0}=\left\{\begin{array}{ll}
\beta & \beta < \beta_c(H)\\
\beta_c(H) & \beta \ge \beta_c(H)
\end{array}\right.$$

\begin{figure}[ht]
\hspace*{3cm}
\begin{picture}(0,0)%
\includegraphics{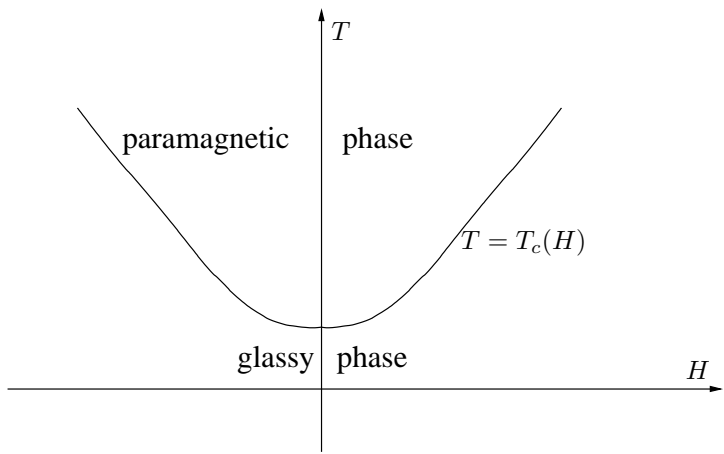}%
\end{picture}%
\setlength{\unitlength}{3947sp}%
\begingroup\makeatletter\ifx\SetFigFont\undefined%
\gdef\SetFigFont#1#2#3#4#5{%
  \reset@font\fontsize{#1}{#2pt}%
  \fontfamily{#3}\fontseries{#4}\fontshape{#5}%
  \selectfont}%
\fi\endgroup%
\begin{picture}(4518,2814)(320,-2968)
\put(2369,-355){\makebox(0,0)[lb]{\smash{\SetFigFont{10}{12.0}{\rmdefault}{\mddefault}{\itdefault}{$T$}%
}}}
\put(4594,-2486){\makebox(0,0)[lb]{\smash{\SetFigFont{10}{12.0}{\rmdefault}{\mddefault}{\itdefault}{$H$}%
}}}
\put(3184,-1671){\makebox(0,0)[lb]{\smash{\SetFigFont{10}{12.0}{\rmdefault}{\mddefault}{\itdefault}{$T=T_{c}(H)$}%
}}}
\end{picture}
\caption{Phase diagram of temperature vs.\ magnetic field.}
\label{gen2}
\end{figure}

\subsection{Joint Source--Channel Code Ensembles and the REM in a Magnetic Field}

In this subsection, we analyze the behavior of a finite--temperature decoder
for a typical randomly selected code using the tools of the analysis of the REM in a magnetic field.
Using the viewpoint of the magnetic properties of the REM,
it will be seen that the source bits play the role of spins in a magnetic field 
whose intensity is $H=\frac{1}{2}\ln\frac{q}{1-q}$, where $q$ is the probability
that $u_i=1$ for each $i$.
Accordingly, instead of the
binary alphabet $\{0,1\}$ that we used before, 
it will prove more convenient to let each $u_i$ assume values in $\{-1,+1\}$.
Another slight change in notation, that will take place mostly for the sake of convenience,
is that instead of defining channel capacity
and coding rates in terms of bits, we will define them in units of {\it nats}, 
where $1$ nat $=\ln 2$ bits. This means that
logarithms will be taken to the natural basis $e$ rather than the base 2. Accordingly,
$h(q)$ will be redefined hereafter as $h(q)=-q\ln q-(1-q)\ln(1-q)$ and
the capacity of the binary channel considered in Section 2 will be redefined as $C=\ln 2-h(q)=
\ln 2+q\ln q+(1-q)\ln(1-q)$.

Consider then a binary memoryless source sequence, $u_1,u_2,\ldots$, $u_i\in\{-1,+1\}$,
with a parameter $q=\mbox{Pr}\{u_i=1\}$ and a binary symmetric
channel with parameter $p$, which as
described in Section 2, is assumed to operate $\theta$ times faster than the source,
in other words, the channel
transmits $\theta$ bits during the time that the source generates one bit.
The number $\theta$ is a positive real which will be assumed fixed throughout the sequel.
Consider a joint source--channel code that receives a source vector of length
$N$, $\bu=(u_1,\ldots,u_N)$, and produces a channel input vector $\bx(\bu)$ of length $n=N\theta$.
The block encoder is generated by random selection: We randomly draw $2^N$ binary
$n$-vectors, $\{\bx(\bu),~\bu\in\{-1,+1\}^{N}\}$, independently, by fair coin tossing.
As described in Section 2, when the input to the 
encoder is $\bu$, the encoder transmits the corresponding
codeword $\bx=\bx(\bu)$, and the
decoder, upon receiving the channel output $\by$, applies 
a finite--temperature decoder
$$\hat{u}_i=\mbox{argmax}_{u\in\{-1,+1\}}\sum_{\bu:~u_i=u}[P(\bu)P(\by|\bx(\bu))]^\beta,~~~
i=1,2,\ldots,N.$$ 
We can think of this decoder as a symbol MAP decoder pertaining to 
a {\it posterior} distribution given by
\begin{eqnarray}
P_\beta(\bu|\by)&=&\frac{[P(\bu)P(\by|\bx(\bu))]^\beta}
{\sum_{\bu'}[P(\bu')P(\by|\bx(\bu'))]^\beta}\nonumber\\
&=&\frac{e^{-\beta\ln[1/P(\bu)P(\by|\bx(\bu))]}}{
\sum_{\bu'}e^{-\beta\ln[1/P(\bu)P(\by|\bx(\bu))]}}
\end{eqnarray}
where in the second line we presented this distribution in the form of 
the Boltzmann--Gibbs distribution with an Hamiltonian given by $\ln[1/P(\bu)P(\by|\bx(\bu))]$
(see also e.g., \cite[Chap.\ 6]{MM06}).
The corresponding {\it partition function} is then
\begin{eqnarray}
Z(\beta)&=& 
\sum_{\bu} \left[P(\bu)P(\by|\bx(\bu))\right]^\beta\nonumber\\
&=&\left[P(\bu_0)P(\by|\bx(\bu_0))\right]^\beta+
\sum_{\bu\ne\bu_0} \left[P(\bu)P(\by|\bx(\bu))\right]^\beta\nonumber\\
&\dfn&Z_c(\beta)+Z_e(\beta),
\end{eqnarray}
where we have separated the partition function into two contributions: $Z_c(\beta)$,
corresponding to the correct source sequence $\bu_0$ that was actually generated
by the source and fed into the encoder, and $Z_e(\beta)$
corresponding to all other possible messages. 
Now, since typically, the source produces sequences with about
$Nq$ occurrences of +1 and $N(1-q)$
occurrences of -1, and the channel flips about $np$ out of $n$ of the transmitted bits,
$Z_c(\beta)$ is typically around $e^{-N\beta h(q)}\cdot e^{-n\beta h(p)}=
e^{-N\beta[h(q)+\theta h(p)]}$. 
On the other hand, 
as we will show now, $Z_e(\beta)$ behaves like
the REM in a magnetic field whose intensity is
$$H=\frac{1}{2}\ln\frac{q}{1-q}.$$
Accordingly, we will henceforth denote $Z_e(\beta)$
also by $Z_e(\beta,H)$, to emphasize the analogy to the REM in a magnetic field.

To see that $Z_e(\beta,H)$ behaves like the REM
in a magnetic field, consider 
the following: first, denote by $N_1(\bu)$ the number of $+1$'s in $\bu$,
so that the magnetization, $m(\bu)\dfn\frac{1}{N}[\sum_{i=1}^N1\{u_i=+1\}-\sum_{i=1}^N1\{u_i=-1\}]$,
pertaining to spin configuration $\bu$,
is given by $m(\bu)=2N_1(\bu)/N-1$. Equivalently, 
$N_1(\bu)=N(1+m(\bu))/2$,
and then
\begin{eqnarray}
P(\bu)&=&q^{N_1(\bu)}(1-q)^{N-N_1(\bu)}\nonumber\\
&=&(1-q)^{N}\left(\frac{q}{1-q}\right)^{N(1+m(\bu))/2}\nonumber\\
&=&[q(1-q)]^{N/2}\left(\frac{q}{1-q}\right)^{Nm(\bu))/2}\nonumber\\
&=&[q(1-q)]^{N/2}e^{Nm(\bu)H}
\end{eqnarray}
where $H$ is defined as above.
By the same token, for the binary symmetric channel we have:
$$P(\by|\bx)=p^{d_H(\bx,\by)}(1-p)^{n-d_H(\bx,\by)}=(1-p)^n e^{-Bd_H(\bx,\by)}$$
where $B=\ln\frac{1-p}{p}$ and $d_H(\bx,\by)$ is the Hamming distance between $\bx$ and $\by$,
namely, the number of places $\{i\}$ where $y_i\ne x_i$.
Thus,
\begin{eqnarray}
Z_e(\beta,H)&=&[q(1-q)]^{N\beta/2}\sum_m\left[\sum_{\bx(\bu):~m(\bu)=m}
e^{-\beta\ln[1/P(\by|\bx(\bu))]}\right]e^{N\beta mH}\nonumber\\
&=&[q(1-q)]^{\beta N/2}(1-p)^{n\beta}\sum_m\left[\sum_{\bx(\bu):~m(\bu)=m}
e^{-\beta Bd_H(\bx(\bu),\by)}\right]e^{\beta NmH}\nonumber\\
&\dfn&[q(1-q)]^{N\beta/2}(1-p)^{n\beta}\sum_m\zeta(\beta,m)
e^{\beta NmH}
\end{eqnarray}
where the resemblance to eq.\ (\ref{rempart}) is self evident,
with $\zeta(\beta,m)$ being redefined as the second bracketed term.
In analogy to the above analysis of the REM, $\zeta(\beta,m)$ here
behaves like in the REM without a magnetic field, namely, it contains exponentially
$e^{Nh((1+m)/2)}=e^{nh((1+m)/2)/\theta}$ terms, with the random energy levels
of the REM being replaced now by random Hamming distances $\{d_H(\bx(\bu),\by)\}$
that are induced by the random selection of the code $\{\bx(\bu)\}$.\footnote{Of course, the
channel output vector $\by$ is also random, but this randomness does not play any essential
role here. This discussion applies as well for every given $\by$.}
Using the same
considerations as with the REM (see also \cite{MM06}), $\zeta(\beta,m)$
can be represented as $\sum_\delta N_{\by,m}(\delta)e^{-\beta Bn\delta}$, where $N_{\by,m}(\delta)$
is the number of vectors $\{\bu\}$ with $m(\bu)=m$ and $d_H(\bx(\bu),\by)=n\delta$.
Since $N_{\by,m}(\delta)$ is the sum of $e^{nh((1+m)/2)/\theta}$ many i.i.d.\ binary random
variables of the form $1\{d_H(\bx(\bu),\by)=n\delta\}$ (again, with randomness induced by the random
selection of $\bx(\bu)$), each with expectation given by $\mbox{Pr}\{d_H(\bx(\bu),\by)=n\delta\}
\exe e^{n[h(\delta)-\ln 2]}$,
then $N_{\by,m}(\delta)$ is typically zero for all $\delta$
such that $h((1+m)/2)/\theta+h(\delta)-\ln 2 < 0$, and is typically
around its expectation, $e^{n[h((1+m)/2)/\theta+h(\delta)-\ln 2]}$,
for all $\delta$ such that $h((1+m)/2)/\theta+h(\delta)-\ln 2\ge 0$.

\begin{figure}[ht]
\hspace*{3cm}
\begin{picture}(0,0)%
\includegraphics{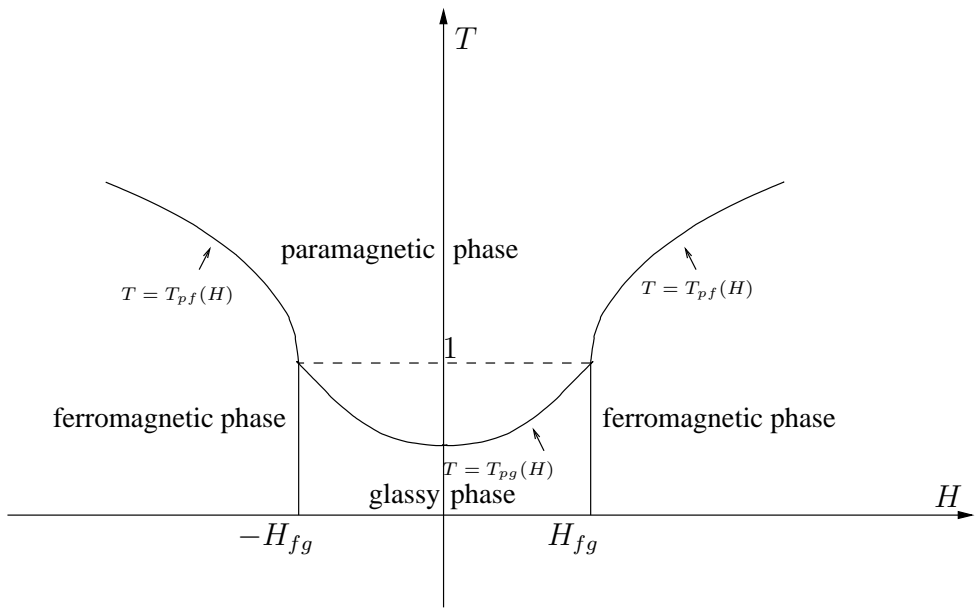}%
\end{picture}%
\setlength{\unitlength}{3947sp}%
\begingroup\makeatletter\ifx\SetFigFont\undefined%
\gdef\SetFigFont#1#2#3#4#5{%
  \reset@font\fontsize{#1}{#2pt}%
  \fontfamily{#3}\fontseries{#4}\fontshape{#5}%
  \selectfont}%
\fi\endgroup%
\begin{picture}(6098,3794)(182,-3958)
\put(3019,-432){\makebox(0,0)[lb]{\smash{\SetFigFont{12}{14.4}{\rmdefault}{\mddefault}{\itdefault}{$T$}%
}}}
\put(6027,-3312){\makebox(0,0)[lb]{\smash{\SetFigFont{12}{14.4}{\rmdefault}{\mddefault}{\itdefault}{$H$}%
}}}
\put(2941,-2374){\makebox(0,0)[lb]{\smash{\SetFigFont{12}{14.4}{\rmdefault}{\mddefault}{\itdefault}{$1$}%
}}}
\put(922,-2008){\makebox(0,0)[lb]{\smash{\SetFigFont{7}{8.4}{\rmdefault}{\mddefault}{\itdefault}{$T=T_{pf}(H)$}%
}}}
\put(1655,-3549){\makebox(0,0)[lb]{\smash{\SetFigFont{12}{14.4}{\rmdefault}{\mddefault}{\itdefault}{$-H_{fg}$}%
}}}
\put(4188,-1971){\makebox(0,0)[lb]{\smash{\SetFigFont{7}{8.4}{\rmdefault}{\mddefault}{\itdefault}{$T=T_{pf}(H)$}%
}}}
\put(2941,-3108){\makebox(0,0)[lb]{\smash{\SetFigFont{7}{8.4}{\rmdefault}{\mddefault}{\itdefault}{$T=T_{pg}(H)$}%
}}}
\put(3597,-3549){\makebox(0,0)[lb]{\smash{\SetFigFont{12}{14.4}{\rmdefault}{\mddefault}{\itdefault}{$H_{fg}$}%
}}}
\end{picture}
\caption{Phase diagram of joint source channel coding: temperature vs.\ magnetic field.}
\label{gen3}
\end{figure}

Defining now the Gilbert--Varshamov distance $\delta_{GV}(R)$ 
\cite[Chap.\ 6]{MM06} as the solution $\delta\le 1/2$
to the equation $h(\delta)=\ln 2-R$, the condition $h((1+m)/2)/\theta+h(\delta)-\ln 2\ge 0$
is equivalent to the condition $\delta_{GV}(h((1+m)/2)/\theta)\le \delta\le 
1-\delta_{GV}(h((1+m)/2)/\theta)$. Thus, for a typical randomly selected code,
\begin{eqnarray}
\phi(\beta,m)&\dfn&\lim_{n\to\infty}\frac{\ln\zeta(\beta,m)}{n}\nonumber\\
&=&\max_{\delta\in[\delta_{GV}(h((1+m)/2)/\theta),1-\delta_{GV}(h((1+m)/2)/\theta]}
\left[\frac{1}{\theta}h\left(\frac{1+m}{2}\right)+h(\delta)-\ln 2-\beta B\delta\right]\nonumber\\
&=&\left\{\begin{array}{ll}
\frac{1}{\theta}h\left(\frac{1+m}{2}\right)-\ln 2 +h(p_\beta)-\beta Bp_\beta & 
p_\beta \ge \delta_{GV}\left(\frac{1}{\theta}h\left(\frac{1+m}{2}\right)\right) \\
-\beta B \delta_{GV}\left(\frac{1}{\theta}h\left(\frac{1+m}{2}\right)\right) &
p_\beta < \delta_{GV}\left(\frac{1}{\theta}h\left(\frac{1+m}{2}\right)\right)
\end{array}\right. 
\end{eqnarray}
where $p_\beta\dfn p^\beta/(p^\beta+(1-p)^\beta)$.
The condition $p_\beta \ge \delta_{GV}(\frac{1}{\theta}h(\frac{1+m}{2}))$
is equivalent to the condition
$$\beta \le \beta_0(m)\dfn
\frac{1}{B}\ln\left[\frac{1-\delta_{GV}(h((1+m)/2)/\theta)}{\delta_{GV}(h((1+m)/2)/\theta)}\right].$$
The exponential order of 
$\sum_m\zeta(\beta,m)e^{N\beta mH}$,
as a function of $N$
is then 
$$\psi(\beta,H)\dfn\lim_{N\to\infty}\frac{1}{N}\ln \left[\sum_m\zeta(\beta,m)e^{N\beta mH}\right]=
\max_m[\theta\phi(\beta,m)+\beta mH].$$
For small enough $\beta$, the dominant
value of $m$ is the one that maximizes $[h((1+m)/2)+\beta mH]$, namely,
the well--known paramagnetic magnetization $m=m_p(\beta,H)=\tanh(\beta H)$.
This is true as long as $\beta\le\beta_0(\tanh(\beta H))$.
Consider then the equation
$$\beta=\beta_0(\tanh(\beta H))$$
where the unknown is $\beta$, or equivalently, the equation
$$\ln 2-h(p_\beta)= \frac{1}{\theta}h\left(\frac{1+\tanh(\beta H)}{2}\right).$$
Now $h((1+\tanh(\beta H))/2)$ is decreasing with $\beta$, 
while $[\ln 2-h(p_\beta)]$ is increasing. At $\beta=0$,
$\ln 2-h(p_\beta)=0$ 
whereas $h((1+\tanh(\beta H))/2)/\theta=\ln 2/\theta$.
As $\beta\to\infty$, $\ln 2-h(p_\beta)\to \ln 2$ whereas $h((1+\tanh(\beta H))/2)/\theta
\to 0$, provided that $H\ne 0$. 
Thus, for $H\ne 0$, there must be a unique solution, which we shall denote by $\beta_{pg}(H)$,
where the subscript ``pg'' stands for the fact that this is the boundary curve
between the paramagnetic phase and the glassy phase.
Since $h((1+\tanh(\beta H))/2))/\theta$ is decreasing with $|H|$,
$\beta_{pg}(H)$ is decreasing in $|H|$, i.e., 
the temperature $T_{pg}(H)=1/\beta_{pg}(H)$ is increasing in $|H|$, as before (see Fig.\ \ref{gen3}).
As for the case $H=0$, for $\theta > 1$, we have
$$\beta_{pg}(0)=\frac{1}{B}\ln\left[\frac{1-h^{-1}((1-1/\theta)\ln 2)}
{h^{-1}((1-1/\theta)\ln 2)}\right].$$
For $0< \theta \le 1$, $\beta_{pg}(0)=\infty$, namely, $T_{pg}(0)=0$, which
means that there is no phase transition as the behavior is paramagnetic at all temperatures.
In the same manner, it is easy to see that $\beta_{pg}(\infty)=0$ for all $\theta > 0$, which
is another case where there are no phase transitions, but this time, it is 
a glassy behavior at all temperatures.

As long as $\beta \le \beta_{pg}(H)$, we have
$$\psi(\beta,H)=\psi_p(\beta,H)\dfn h\left(\frac{1+\tanh(\beta H)}{2}\right)-
\theta(\ln 2-h(p_\beta)+
\beta Bp_\beta)+\beta H\tanh(\beta H).$$
On the other hand, for $\beta > \beta_{pg}(H)$, the system is in the glassy phase.
In this case, 
$$\psi(\beta,H)=\psi_g(\beta,H)\dfn
\beta\max_m\left[mH-
\theta B\delta_{GV}\left(\frac{1}{\theta}h\left(\frac{1+m}{2}\right)\right)\right]$$
thus, the maximizing $m$ depends only on $H$ but not on $\beta$. In this case,
we have $m=m_g(H)=\tanh(\beta_{pg}(H)\cdot H)$ and so
\begin{eqnarray}
\psi_g(\beta,H)&=&\beta\left[H\tanh(\beta_{pg}(H)\cdot H)-
B\theta\delta_{GV}\left(\frac{1}{\theta}h\left(\frac{1+
\tanh(\beta_{pg}(H)\cdot H)}{2}\right)\right)\right]\nonumber\\
&=&\beta\left[H\tanh(\beta_{pg}(H)\cdot H)-
B\theta p_{\beta_{pg}(H)}\right].
\end{eqnarray}
The free--energy density associated with erroneous messages is therefore given by
$$F_e(\beta,H)\dfn-\lim_{N\to\infty}\frac{\ln Z_e(\beta,H)}{N\beta}=
-\frac{1}{2}\ln[q(1-q)]-\theta\ln(1-p)-\frac{\psi(\beta,H)}{\beta}$$
i.e.,
$$F_e(\beta,H)=\left\{\begin{array}{ll}
F_p(\beta,H) & \beta \le \beta_{pg}(H)\\
F_g(H) & \beta > \beta_{pg}(H) \end{array}\right.$$
where
$$F_p(\beta,H)=
-\frac{1}{2}\ln[q(1-q)]-\theta\ln(1-p)-\frac{1}{\beta}\left[
h\left(\frac{1+\tanh(\beta H)}{2}\right)-\theta(\ln 2-h(p_\beta))\right]
+\theta Bp_\beta-H\tanh(\beta H)$$
and
$$F_g(H)=
-\frac{1}{2}\ln[q(1-q)]-\theta\ln(1-p)-
H\tanh(\beta_{pg}(H)\cdot H)+
B\theta p_{\beta_{pg}(H)}.$$

The boundary between the ferromagnetic phase (where $Z_c(\beta)$ is 
the dominant term in $Z(\beta)$) and the glassy phase 
is the vertical line (see Fig.\ \ref{gen3})
$H=H_{fg}$, where $H_{fg}$ is the solution to the equation
$$\frac{h(q)}{\theta}+h(p)= -\frac{1}{2\theta}\ln[q(1-q)]-\ln(1-p)-
\frac{H\tanh(\beta_{pg}(H)H)}{\theta}+Bp_{\beta_{pg}(H)}$$
which after rearranging terms becomes
$$Bp-\frac{H\tanh(H)}{\theta}=Bp_{\beta_{pg}(H)}-\frac{H\tanh(\beta_{pg}(H)\cdot H)}{\theta},$$
whose solution in turn is achieved when $\beta_{pg}(H)=1$, i.e.,
$$p=\delta_{GV}\left(\frac{1}{\theta}
h\left(\frac{1+\tanh(H)}{2}\right)\right)\equiv \delta_{GV}\left(\frac{h(q)}{\theta}\right),$$
which is nothing but the boundary of reliable communication (\ref{relcond}).
Thus, 
$$H_{fg}=\frac{1}{2}\ln \frac{q^*}{1-q^*}~~
\mbox{with}~~q^*=1-h^{-1}(\theta(\ln 2-h(p)),$$
where $h^{-1}(\cdot)$ is the inverse of the function $h(\cdot)$ in the range
where the argument is in $[0,\frac{1}{2}]$.
The vertical line $H=H_{fg}$ intersects the paramagnetic--glassy boundary curve $T=T_{pg}(H)$
at the triple point $(H,T)=(H_{fg},1)$,
namely, $T_{pg}(H_{fg})=1$. The ferromagnetic region,
pertaining to correct decoding (where $m=2q-1=\tanh(H)$),
is $\{(H,T):~|H|\ge H_{fg},~T<T_{pf}(H)\}$,
where $T=T_{pf}(H)$ is paramagnetic--ferromagnetic boundary curve 
(see Fig.\ \ref{gen3}) given by the solution $\beta=1/T$
of the equation
$$\beta p B-\frac{\beta H\tanh(H)}{\theta}=\ln 2+\beta p_\beta B-h(p_\beta)-
\frac{1}{\theta}h\left(\frac{1+\tanh(\beta H)}{2}
\right)-\frac{\beta}{\theta} H\tanh(\beta H)$$
for every given $H$ which is larger than $H_{fg}$ in absolute value.
As can be seen, it also contains the point $(H,T)=(H_{fg},1)$.

\vspace{0.5cm}

\noindent
{\bf Discussion:}
We see that correct decoding occurs in a sufficiently strong magnetic field.
This is not surprising as a strong magnetic field corresponds to a low--entropy source
which can be transmitted reliably.
The above exposition of the magnetization as a function of $H$ and $T$
is instructive for the understanding of typical error patterns in joint source--channel coding.
At very low temperatures (like in word MAP decoding, which corresponds to $\beta\to\infty$), 
the (sub--exponentially few) typical patterns of the
erroneneously decoded vectors $\{\bu\}$ have magnetization dictated by the frozen 
phase, namely, $m_g(H)=\tanh(\beta_{pg}(H)\cdot H)$, independently of the decoding temperature.
For magnetic fields smaller than $H_{fg}$ in absolute value (namely, for sources with
high entropy), $\beta_{pg}(H) > 1$, which means that the magnetization of a typical
erroneously decoded sequence is {\it higher} than that of a typical (correct) source sequence
which is $m_f=2q-1=\tanh(H)$. If the working temperature is lower than $T_{pg}(0)$, this remains true
no matter how small $|H|$ is. If, on the other hand, $T_{pg}(0) < T < 1$, then when the magnetic
field is reduced, the magnetization of the (exponentially many) erroneously decoded vectors $\{\bu\}$
is given by $m_p(\beta,H)=\tanh(\beta H)$, which is still higher than that of the typical
source vector $\bu$, but now it is temperature--dependent.

\section{Ensemble Performance of Codes and Free Energies}

In this section, we provide bounds on the ensemble performance of joint
source channel codes for the binary symmetric source and the
binary symmetric channel. In particular, we examine the exponential decay 
rate of the average probability of correct decoding (the correct decoding exponent, for short)
when the condition for reliable communication (\ref{relcond}) is violated as well as
the exponential decay rate of the
average probability of error (error exponent) when this condition holds. As will be seen, the former
is intimately related to the free energy in the glassy phase, whereas the latter is
strongly related to the free energy in the paramagnetic phase.

The relationship between the previous derivations and
both the correct decoding exponent and the
error exponent stems from the fact both performance measures
are bounded by expressions that are strongly related to the
partition function $Z_e(\beta)$.

\subsection{The Correct Decoding Exponent}

The probability of correct decoding 
pertaining to the word MAP decoder is well known (and can easily be shown) to be given by
\begin{eqnarray}
P_c&=&\sum_{\by}\max_{\bu}[P(\bu)P(\by|\bx(\bu))]\nonumber\\
&=&\sum_{\by}\lim_{\beta\to\infty}\left[\sum_{\bu}
P^\beta(\bu)P^\beta(\by|\bx(\bu))\right]^{1/\beta}\nonumber\\
&=&[q(1-q)]^{N/2}(1-p)^n\sum_{\by}
\lim_{\beta\to\infty}\left[\sum_m\zeta(\beta,m)e^{N\beta mH}\right]^{1/\beta}\nonumber\\
&\exe&[q(1-q)]^{N/2}(1-p)^n\sum_{\by}
\lim_{\beta\to\infty}\sum_m\zeta^{1/\beta}(\beta,m)e^{NmH},
\end{eqnarray}
where with a slight abuse of notation, here $\zeta(\beta,m)$ is redefined to include
{\it all} messages $\{\bu\}$, including the correct one. Now, taking the ensemble average:
$$\bar{P}_c\exe [q(1-q)]^{N/2}(1-p)^n 
\sum_{\by}\lim_{\beta\to\infty}\sum_m\bE\{\zeta^{1/\beta}(\beta,m)\}\cdot e^{NmH}.$$
Now,
\begin{eqnarray}
\bE\{\zeta^{1/\beta}(\beta,m)\}&=&\bE\left\{\left[
\sum_{\delta}N_{\by,m}(\delta)e^{-\beta Bn\delta}\right]^{1/\beta}\right\}\nonumber\\
&\exe&\bE\left\{\left[
\max_{\delta}N_{\by,m}(\delta)e^{-\beta Bn\delta}\right]^{1/\beta}\right\}\nonumber\\
&=&\bE\left\{
\max_{\delta}N_{\by,m}^{1/\beta}(\delta)e^{- Bn\delta}\right\}\nonumber\\
&\exe&\sum_{\delta}\bE\{N_{\by,m}^{1/\beta}(\delta)\}\cdot e^{-Bn\delta}
\end{eqnarray}
where again,  $N_{\by,m}(\delta)$ is the number of codewords 
$\{\bx(\bu)\}$, corresponding to source words with $m(\bu)=m$,
which fall at Hamming distance $n\delta$ from $\by$.
Now, as shown in \cite{Merhav07},\cite[Appendix]{Merhav07a},
\begin{equation}
\bE\{N_{\by,m}^{1/\beta}(\delta)\}\exe\left\{\begin{array}{ll}
\exp\left\{n\left[\frac{1}{\theta}h\left(\frac{1+m}{2}\right)+h(\delta)-\ln 2\right]\right\} & 
\frac{1}{\theta}h\left(\frac{1+m}{2}\right)+h(\delta)<\ln 2\\
\exp\left\{n\left[\frac{1}{\theta}h\left(\frac{1+m}{2}\right)+h(\delta)-\ln 2\right]/\beta\right\} & 
\frac{1}{\theta}h\left(\frac{1+m}{2}\right)+h(\delta)\ge\ln 2\end{array}\right.
\end{equation}
Thus,
\begin{equation}
\lim_{\beta\to\infty}\bE\{\zeta^{1/\beta}(\beta,m)\}\exe
\exp\left\{n\left[\frac{1}{\theta}h\left(\frac{1+m}{2}\right)-\ln 2+
\max_{\delta\le\delta_{GV}(h((1+m)/2)/\theta)}\{h(\delta)-B\delta\}\right]\right\}
\end{equation}
and so,
\begin{eqnarray}
\bar{P}_c&\exe&[q(1-q)]^{N/2}(1-p)^n \sum_{\by}\sum_m \left[
\exp\left\{n\left[\frac{1}{\theta}h\left(\frac{1+m}{2}\right)-\ln 2+\right.\right.\right.\nonumber\\
& &\left.\left.\left.
\max_{\delta\le\delta_{GV}(h((1+m)/2)/\theta)}\{h(\delta)-B\delta\}\right]\right\}\right]e^{NmH}.
\end{eqnarray}
The dominant $m$ is the one that maximizes
$$\frac{1}{\theta}h\left(\frac{1+m}{2}\right)-
\ln 2+\max_{\delta\le\delta_{GV}(h((1+m)/2)/\theta)}\{h(\delta)-B\delta\}+\frac{mH}{\theta}$$
Now, if
$h(p) \le \ln 2-h((1+m)/2)/\theta$, 
then the inner maximization is attained at $\delta=p$ and we get
$$\frac{1}{\theta}h\left(\frac{1+\tanh(H)}{2}\right)-\ln 2+h(p)-Bp+\frac{H\tanh(H)}{\theta}.$$
If $h(p) \le \ln 2-h((1+\tanh(H))/2)/\theta$, namely, the condition for reliable communication holds,
this indeed happens. In this case, we get
$$\bar{P}_c\exe [q(1-q)]^{N/2}(1-p)^n\cdot 2^n\cdot 
\exp\left\{n\left[\frac{h(q)}{\theta}-\ln 2+h(p)-Bp+\frac{2q-1}{2\theta}\ln
\frac{q}{1-q}\right]\right\}=1,$$
as expected.
Otherwise, the maximum is attained at the boundary of the allowed range of $\delta$, and we get
$$\max_m\left[\frac{mH}{\theta}-B\delta_{GV}\left(\frac{1}{\theta}
h\left(\frac{1+m}{2}\right)\right)\right]
=\frac{H}{\theta}\tanh(\beta_{pg}(H)\cdot H)-
B\delta_{GV}\left(\frac{1}{\theta}h\left(\frac{1+\tanh(\beta_{pg}(H)\cdot H}{2}\right)\right)$$
and so, the correct decoding exponent is 
\begin{eqnarray}
E_c&\dfn&-\lim_{N\to\infty}\frac{\ln\bar{P}_c}{N}\nonumber\\
&=&-\frac{1}{2}\ln[q(1-q)]-\theta\ln[2(1-p)]+\nonumber\\
& &B\theta\delta_{GV}
\left(\frac{1}{\theta}h\left(\frac{1+
\tanh(\beta_{pg}(H)\cdot H}{2}\right)\right)-
H\tanh(\beta_{pg}(H)\cdot H)\nonumber\\
&=&F_g(H)-\theta\ln 2.
\end{eqnarray}
Thus, we obtained a very simple relationship between the
correct decoding exponent and the glassy free energy.
The ferromagnetic--glassy phase transition is 
exactly the transition from $E_c=0$ to $E_c > 0$.
The dominant magnetization of the correct 
decoding event is then $m_g(H)=\tanh(\beta_{pg}(H)\cdot H)$, i.e.,
the dominant
(rare) event of correct decoding is when the source vector $\bu$ has
the (non--typical) magentization $m_g(H)$. If the condition of reliable communication
does not hold, i.e., $h(q)/\theta > \ln 2 -h(p)$, then the word MAP decoder ($\beta\to\infty$)
works in the glassy regime, but the symbol MAP decoder ($\beta=1$) works in the paramagnetic
regime. The computation of $\bar{P}_c$ for the word MAP decoder is carried out also in the
glassy regime.

\subsection{The Error Exponent}

We begin by using Gallager's techniques (see 
\cite[Problem 5.16, pp.\ 534--535]{Gallager68}):
The probability of error for a given code and the word MAP decoder is given by
$$P_e=\sum_{\bu}P(\bu)\sum_{\by}P(\by|\bx(\bu))\cdot 1\{\exists~\bu':~P(\bu')P(\by|\bx(\bu'))\ge
P(\bu)P(\by|\bx(\bu))\}.$$
Now, it is easy to see that whenever an error occurs
$$\sum_{\bu'\ne \bu}
\left[\frac{P(\bu')P(\by|\bx(\bu'))}{P(\bu)P(\by|\bx(\bu))}\right]^\beta\ge 1.$$
for every $\beta\ge 0$. Thus,
$$1\{\exists~\bu':~P(\bu')P(\by|\bx(\bu'))\ge
P(\bu)P(\by|\bx(\bu))\}\le\left(\sum_{\bu'\ne\bu}\left[\frac{P(\bu')P(\by|\bx(\bu'))}
{P(\bu)P(\by|\bx(\bu))}\right]^\beta\right)^\rho$$
for every $\rho\ge 0$. Substituting the right--hand side into the expression of $P_e$,
we get the following upper bound:
\begin{equation}
P_e\le\sum_{\bu}P(\bu)^{1-\rho\beta}\sum_{\by} P(\by|\bx(\bu))^{1-\rho\beta}\left(\sum_{\bu'\ne\bu}
[P(\bu')P(\by|\bx(\bu'))]^\beta\right)^\rho~~~~~\beta\ge 0,~\rho\ge 0.
\end{equation}
Thus, the average error probability over the ensemble of codes is bounded by
\begin{equation}
\bar{P}_e\le \sum_{\bu}P(\bu)^{1-\rho\beta}\sum_{\by}\bE\left\{P(\by|\bX(\bu))^{1-\rho\beta}\right\}\cdot
\bE\left\{\left(\sum_{\bu'\ne\bu}[P(\bu')P(\by|\bX(\bu'))]^\beta\right)^\rho\right\}.
\end{equation}
In the binary symmetric case considered here, the first expectation is given by:
\begin{eqnarray}
\bE\left\{P(\by|\bX(\bu))^{1-\rho\beta}\right\}&=&2^{-n}
\sum_{\bx}\prod_{i=1}^n P(y_i|x_i)^{1-\rho\beta}\nonumber\\
&=&2^{-n}[p^{1-\rho\beta}+(1-p)^{1-\rho\beta}]^n\nonumber\\
&=& e^{-n\gamma(1-\rho\beta)}
\end{eqnarray}
where $\gamma(s)\dfn\ln 2-\ln[p^s+(1-p)^s]$.
The second expectation is handled as follows.
Using the above derived  relation:
$$\sum_{\bu'}[P(\bu')P(\by|\bX(\bu'))]^\beta=
[q(1-q)]^{N\beta/2}(1-p)^{n\beta}\sum_m\zeta(\beta,m)
e^{\beta NmH},$$
we get
\begin{eqnarray}
\left(\sum_{\bu'\ne\bu}[P(\bu')P(\by|\bX(\bu'))]^\beta\right)^\rho&=&
\left([q(1-q)]^{N\beta/2}(1-p)^{n\beta}\sum_m\zeta(\beta,m)e^{\beta NmH}\right)^\rho\nonumber\\
&\exe&[q(1-q)]^{N\beta\rho/2}(1-p)^{n\beta\rho}\sum_m\zeta^\rho(\beta,m)e^{\beta\rho NmH}
\end{eqnarray}
and so, assuming $\rho\in[0,1]$, and using Jensen's inequality
\begin{eqnarray}
& &\bE\left\{\left(\sum_{\bu'\ne\bu}[P(\bu')
P(\by|\bX(\bu'))]^\beta\right)^\rho\right\}\nonumber\\
&\le&[q(1-q)]^{N\beta\rho/2}(1-p)^{n\beta\rho}
\sum_m[\bE\{\zeta(\beta,m)\}]^\rho e^{\beta\rho mHN}\nonumber\\
&\exe&[q(1-q)]^{N\beta\rho/2}(1-p)^{n\beta\rho}
\sum_m\sum_\delta\exp\left\{n\rho\left[
\frac{1}{\theta}h\left(\frac{1+m}{2}\right)+h(\delta)-\ln 2-
\beta B\delta\right]\right\}\cdot
e^{\beta\rho mHN}\nonumber\\
&\exe&[q(1-q)]^{N\beta\rho/2}(1-p)^{n\beta\rho}\times\nonumber\\
& &\exp\left\{n\rho\left[
\frac{1}{\theta}h\left(\frac{1+\tanh(\beta H)}{2}\right)+h(p_\beta)-\ln 2-\beta Bp_\beta
+\frac{\beta H}{\theta}\tanh(\beta H)\right]\right\}
\end{eqnarray}
We see that the magnetization that dominates the Gallager bound is the
paramagnetic magnetization. 
By plugging this expression back into the bound on $\bar{P}_e$, we get
the error exponent:
\begin{eqnarray}
E&\dfn&-\lim_{N\to\infty}\frac{\ln \bar{P}_e}{N}\nonumber\\
&\ge&-\ln [q^{1-\rho\beta}+(1-q)^{1-\rho\beta}]+\theta[\gamma(1-\rho\beta)-\ln 2]
-\frac{\beta\rho}{2}\ln[q(1-q)]-\beta\rho\theta\ln(1-p)-\nonumber\\
& &\rho\left[h\left(\frac{1+\tanh(\beta H)}{2}\right)+\theta[h(p_\beta)-\ln 2-
\beta Bp_\beta]
+\beta H\tanh(\beta H)\right]\nonumber\\
&=&-\ln [q^{1-\rho\beta}+(1-q)^{1-\rho\beta}]+
\theta[\gamma(1-\rho\beta)-\ln 2]+\rho\beta F_p(\beta,H)\nonumber\\
&=&-\ln \{[p^{1-\rho\beta}+(1-p)^{1-\rho\beta}]^\theta\cdot[q^{1-\rho\beta}+(1-q)^{1-\rho\beta}]\}
+\rho\beta F_p(\beta,H)
\end{eqnarray}
Here, unlike in the computation of the correct decoding exponent, there is a mismatch
between the phase in the $H-T$ plane at which the decoder operatively works, and the
phase at which $\bar{P}_e$ is analyzed: While the former is ferromagnetic, the latter
is paramagnetic regardless of the temperature.

\end{document}